\documentclass[twocolumn,amsmath,amssymb]{revtex4}

\usepackage{amsmath} %
\usepackage{graphicx}
\usepackage{amsfonts}
\usepackage{dcolumn}
\usepackage{bm}

\begin{document}
\title{Large-scale Epitaxial Growth Kinetics of Graphene: A Kinetic Monte Carlo Study}

\author{Huijun Jiang}
\author{Zhonghuai Hou}
\thanks{Corresponding author. E-mail: hzhlj@ustc.edu.cn}
\affiliation{Department of Chemical Physics \& Hefei National Laboratory for Physical Sciences at Microscales, University of Science and Technology of China, Hefei, Anhui 230026, China}
\date{\today}

\begin{abstract}
Epitaxial growth via chemical vapor deposition is considered to be the most promising way towards synthesizing large area graphene with high quality. However, it remains a big theoretical challenge to reveal growth kinetics with atomically energetic and large-scale spatial information included. Here, we propose a minimal kinetic Monte Carlo model to address such an issue on an active catalyst surface with graphene/substrate lattice mismatch, which facilitates us to perform large scale simulations of the growth kinetics over two dimensional surface with growth fronts of complex shapes.  A geometry-determined large-scale growth mechanism is revealed, where the rate-dominating event is found to be $C_{1}$-attachment for concave growth-front segments and $C_{5}$-attachment for others. This growth mechanism leads to an interesting time-resolved growth behavior which is well consistent with that observed in a recent scanning tunneling microscopy experiment.
\end{abstract}
\maketitle

\section{Introduction}
Graphene, a material with outstanding electronic, mechanical, thermal, and optical performance, has gained explosive growth of research interests since the works pioneered by \citeauthor{Sci04000666}\cite{Sci04000666} and by \citeauthor{JPCB04019912}\cite{JPCB04019912} Among several ways to produce graphene\cite{Sci04000666,CSR10000228}, epitaxial growth via chemical vapor deposition (CVD) on metal surfaces is considered to be the most promising one towards synthesizing high-quality and large area graphene at relatively low cost\cite{NtM08000406,NaL08000565,Nat09000706,ACS11003385,Sci09001312}. In order to ensure controllable growth of high quality graphene, great efforts have been contributed to understand the underlying mechanism.\cite{PhysRept14000195,PRB07075429,NaL08000030,PRL08107602,PRL09056808,PRL09166101,NanoLett09004268,ChemMater11001441,NJP08093026,ACSNano13007028}.
Many experimental tools are able to provide the structure or growth information for graphene, such as the scanning tunneling microscopy, low-energy and photoemission microscopy, surface-sensitive electron diffraction, and various growth behaviors have been revealed. Typically, graphene growth is considered to be a precipitate of carbon atoms dissolved in the substrate on transition-metals such as Ni, Co and Fe, and carbon solubility is a key factor for graphene growth\cite{NanoLett09004268}. On Cu surface, due to its low solubility and relative low catalyst activity, growth process is mainly determined by diffusion of surface carbon species\cite{ChemMater11001441}. For active catalyst surfaces such as Ir, Rh and Ru, interaction between carbon and substrate atoms is strong and the graphene/substrate lattices are not well matched,  so that the grown graphene spreads on the surface just like a carpet\cite{NaL08000565} and forms moir\'{e} patterns which are usually super unit cells consisting of hundreds of carbon atoms\cite{NJP08043033,PRL09056808,PRB09085430}. Growth of graphene on these surfaces needs highly supersaturated carbon monomers and the growth rate shows strongly nonlinear dependence on the concentration of monomer\cite{NJP08093026}.

In spite of plenty of experimental achievements, there are yet rare theoretical studies which are surely demanded to provide deep understanding of the growth kinetics during the CVD process. Generally, different theoretical methods can be applied to study the growth-related behaviors at different scales. At the atomic level, first principle calculations
can provide detailed energetic information of the growth events without any empirical parameters. For example, monomer ($C_{1}$) attachment to armchair edges of graphene is energetically favorable, producing grown islands with atomically sharp zigzag edges\cite{AcsNano11009154}.  Density functional theory calculations showed that lattice mismatch between graphene and Ir substrate results that $C_1$ attachment has to overcome a higher energetic barrier for some growth sites than others\cite{JAC12006045}. Although first principle calculations are able to provide detailed information at the atomic scales,  they are usually too computationally expensive to study the growth kinetics. On the other hand,  a rate theory and its refinement at the macroscopic level have been developed to produce a quantitative account of the measured time-dependent carbon adatom density by assuming that graphene islands grow homogeneously via the attachment of five-atom carbon clusters ($C_{5}$)\cite{NaL11002092,JPCM14185008}. With optimized kinetic parameters,  it predicts that the smallest stable precursor to graphene growth is an immobile island composed of six $C_{5}$ clusters. Note that this model is macroscopic and the kinetic parameters used are experiential, such that revealing underlying mechanisms at atomic level is out of its range.

To bridge the gap between microscopic growth events and macroscopic growth kinetics, very recently, we have proposed a multiscale ``standing-on-the-front'' kinetic Monte Carlo (SOF-kMC) approach combining with first principle calculations to study graphene growth on Ir surface\cite{JAC12006045,PRB13054304}. Therein, a quasi-1D model was established by focusing on detailed growth events of carbon species on the growth front to understand the special growth behavior where the average growth rate is a highly nonlinear function of $C_{1}$ concentration\cite{NJP08093026}. Nevertheless, for general growth processes in experiments, there are a plenty of new features beyond the quasi-1D growth kinetics, which are out of the scope of SOF-kMC.  For example, when several adjacent fragments merge into one whole piece of grown graphene, it has to grow over a vacancy island (VI) which is an island shape of empty region inside the flake of graphene. In addition, interesting time-resolved growth behaviors involving VI over the large scale of  moir\'{e} patterns have been reported recently\cite{ACSNano13007028}. Therefore, an applicable approach for full 2D kinetics of large-scale graphene growth is very much desired to obtain a deep understanding of the underlying mechanism, which, however, still remains a big theoretical challenge.

In this paper, we establish a minimal kMC model to address such an issue on an active catalyst surface with lattice mismatch. As shown in the inset of Fig.\ref{fig:180ts}(a), the model is based on a two-type-site kMC lattice where one type atop the underlying substrate atoms (D-site) is difficult for $C_{1}$ attachment and the other (E-site) is easy\cite{JAC12006045}, and consists of only several essential growth events dominating the growth process while the contribution of others are implicitly considered in effective kinetic parameters of these dominating events. With kinetic parameters extracted conveniently from detailed atomic calculations, graphene growth over VI is investigated as an application of the minimal model. Remarkably, we show that the complicated growth behavior observed in experiments can be well reproduced. Detailed analysis reveals that the interesting time dependence of the VI area  is due to the fact that the rate-dominating events for growth is strongly dependent on the local shape of the growth front: The growth event over D-sites can be $C_{1}$-attachment for concave segments while has to be $C_{5}$- attachment for others. Such observation suggests that geometry of the growth front plays important and subtle roles during the large-scale growth process of graphene.

\section{The minimal Model for Large-scale Growth Kinetics}
Generally, a whole process of graphene growth includes a nucleation stage and an epitaxial growth stage after nucleation, both of which are challenges for theoretical studies. Kinetic Monte Carlo (kMC) simulation is good choice for the later challenge for several reasons \cite{NaL11002092}. Firstly, kMC goes beyond the atomistic detail and considers kinetic processes in a time/space coarse-grained manner, which facilitates simulations of growth kinetics with atomic details. Secondly, wealth of spatial information is contained in snapshots of kMC simulation, which can be directly related and compared to the experimental scanning tunneling microscopy images. Nevertheless, as pointed out by \citeauthor{PhysRept14000195}, building up of practicable kMC models for epitaxial growth of graphene is a remarkably complicated problem\cite{PhysRept14000195}. The main issue is about how to capture all the essential events important for growth kinetics while algorithm realization and computational efficiency can be ensured. For large-scale growth of graphene, a full version of kMC model should include all the atomic events related to absorption/desorption and dissociation of carbon sources on metal surface, reactions and diffusion of carbon clusters, attachment/detachment of carbon species to graphene edges, etc. Nevertheless, such a full-kMC is not applicable due to the disparate rates of different kMC events and the entanglement of detailed growth events and the large spatial scale of graphene. In order to overcome the first problem, we have built up the multiscale SOF-kMC as described in Ref. \citenum{JAC12006045} and \citenum{PRB13054304}. To make the second problem tractable, one should identify the very key events that are most relevant to the large-scale growth kinetics.

\begin{figure}[h]
\begin{centering}
\includegraphics[width=0.9\columnwidth]{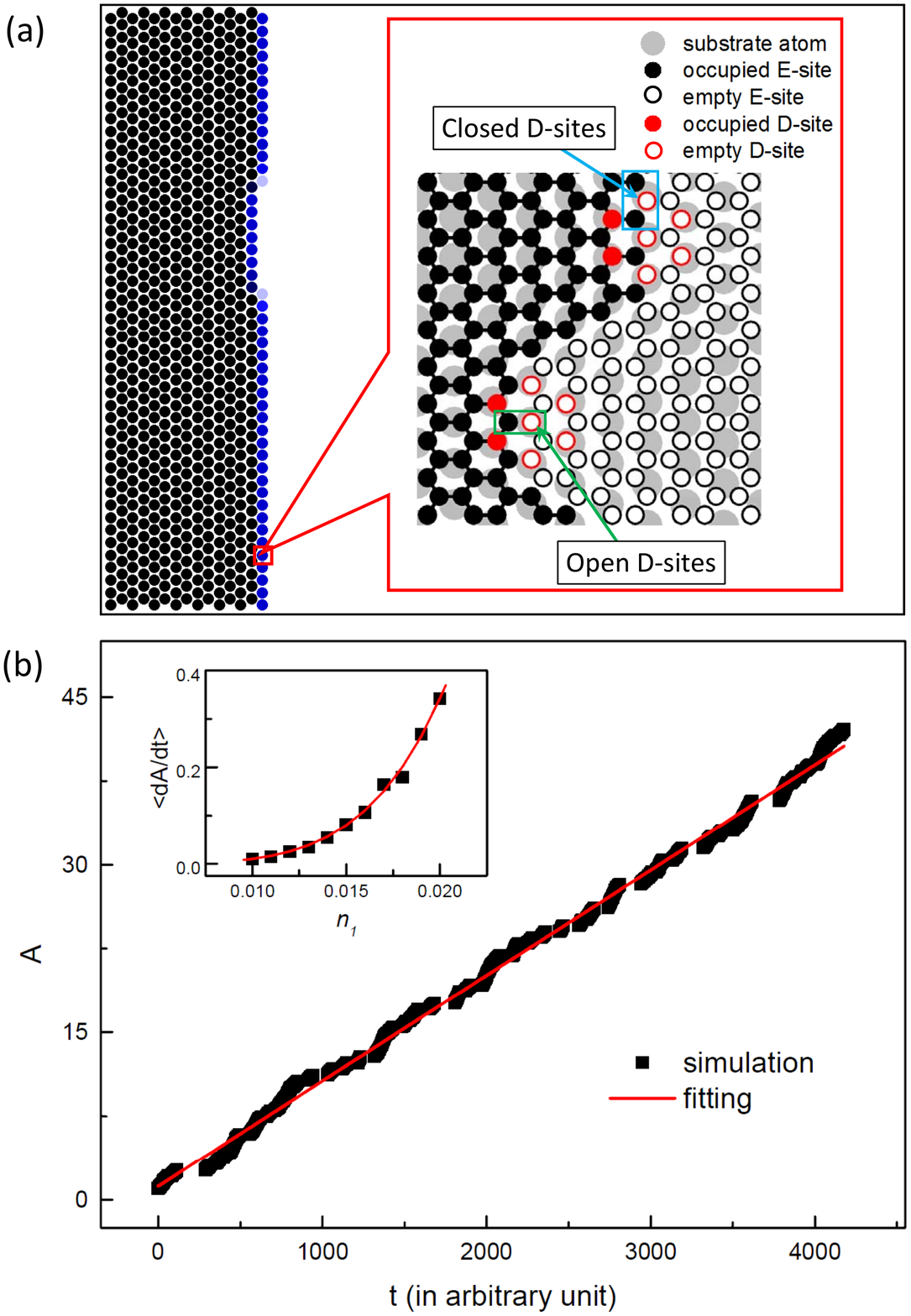}
\par\end{centering}
\protect\caption{(Color online) (a) A typical snapshot during the simulation of large-scale graphene growth with edges straight at the moir\'{e} pattern scale. Each dot represents a moir\'{e} pattern consisting of $10\times20$ growth sites. Black/blue dots refer to fully/partially occupied moir\'{e} patterns, respectively. The inset shows a zoom in of a moir\'{e} pattern, where each circle refers to a growth site. The substrate atoms are in light gray, and E-sites and D-sites are colored by black and red, while solid and open cycles denote occupied and empty sites. The blue/green rectangle encloses a growth fragment with closed/open D-sites, respectively. (b) Linear dependence of area $A$ (normalized by its initial value) for monomer concentration $n_{1}=0.01$ monolayer obtained by using the minimal model (Eq.(\ref{eq:kinetics})). The inset presents the highly nonlinear dependence of average growth rate $\left\langle dA/dt\right\rangle $ on $n_{1}$, which can be well fitted by $\left\langle dA/dt\right\rangle =a+bn_{1}^{\gamma}$ with an exponent $\gamma\approx5$.}
\label{fig:180ts}
\end{figure}

As mentioned above, there are essentially two types of sites over an active catalyst surface such as Ir as shown in the inset of Fig.\ref{fig:180ts}(a), i.e., E-site and D-site, as a consequence of lattice mismatch. As a prior step, one should identify possible elementary events for kMC simulations of large scale growth kinetics. By dividing the whole surface lattice into four regions: The grown graphene sheet, growth front, diffusion layer, and far field, the SOF-kMC is adopted to this end with all the attachment and detachment events of carbon species from $C_1$ up to $C_6$ while all other events in the diffusion layer and far field are compacted into effective carbon fluxes\cite{JAC12006045,PRB13054304}. An important observation is that, while the growth over E-sites is mainly by $C_{1}$ attachment, that over D-sites is a little more complicated \cite{PRB13054304}. As shown in the inset of Fig.\ref{fig:180ts}(a), a moir\'{e} pattern consists of two classes of D-sites along a given zigzag growth front. For a configuration similar to the one enclosed by the blue rectangle, the two E-sites adjacent to the completed zigzag graphene edge have been occupied by $C_1$ very fast, then, attachment of $C_1$ on the D-site will locally close a zigzag graphene edge which is stable and hard to be detached. Thus, such a D-site can be named as a closed D-site. On contrary, for the bottom-left configuration surrounded by the green rectangle, D-site is adjacent directly to the completed zigzag grown front, where the attachment of $C_1$ will result in an open zigzag growth front which is very unstable and easy to be detached. Consequently, this D-site is denoted as an open D-site. Due to these observations, the overall growth rate is found to be dominated by $C_{5}$ and contributed slightly by $C_{6}$ attachment on open D-sites, resulting in a highly nonlinear dependence of growth rate on $C_{1}$ concentration with an exponent slightly bigger than 5\cite{PRB13054304}, while $C_{1}$ attachment on closed D-sites also provides indispensable net contributions to grown graphene.

Based on the above picture, it is convenient for us to refine all the attachment and detachment of carbon species on the growth front to be three key events: $C_{1}$ attachment on E-sites, $C_1$ attachment on closed D-sites, and $C_{5}$ attachment on open D-sites. This leads to a minimal model given by

\begin{equation}
\begin{aligned}C_{1} & \xrightarrow[E]{k_{a}}\: G\\
C_{1} & \xrightarrow[D_{closed}]{k_{b}}\: G\\
C_{5} & \xrightarrow[D_{open}]{k_{c}}\: G
\end{aligned}
,\label{eq:kinetics}
\end{equation}
where $E$, $D_{open}$ and $D_{closed}$ denote E-sites, open D-sites and closed D-sites, respectively, and $G$ refers to the grown graphene.

To accomplish our model, we need to calculate the effective rate constants $k_{a}$, $k_{b}$ and $k_{c}$. Each $k_i (i=a,b,c)$ contains three components, i.e. $k_i\sim n_iexp\{-\epsilon_i/(k_BT)\}f_{0,i}$,  wherein $n_i$ is the surface concentration of the species-i, $\epsilon_i$ is the associated energetic parameters with contributions from the diffusion and attachment processes. Notice that the minimal model only considers the three dominant attaching events as shown in Eq.(\ref{eq:kinetics}), in real growth process nevertheless, other events should also take effects on the growth process. For example, the attached carbon atoms may be detached via detaching events, which leads to a decreasing of the effective attaching rate. To take this into account, a prefactor $f_{0,i}$ denoting a net contribution of event $i$ to the front growth is added as the third component of the effective growth rate constant. As already described in detail in our previous studies\cite{JAC12006045,PRB13054304}, the energetic parameter associated with $\epsilon_i$ can be calculated by first principle calculations and $n_i$ can be derived by assuming a quasi-equilibrium between differently sized carbon species. The factor $f_{0,i}$ can be extracted from detailed simulations with full attachment and detachment events on a small sized lattice as shown in Ref.\citenum{PRB13054304}.

Since all other events, such as combining, diffusion and detachment of carbon species, have been compacted in the effective rate constant $k_i$, a standard kMC simulation is sufficient for the simulation of large scale growth of graphene. One only needs to figure out configurations on the growth front ready for attachment, and the rate for each event is $r_i=k_iN_i$ where $N_i$ is the number of configuration for event $i$. To check the validity of our approach, we first apply the above minimal model to investigate graphene growth with steady carbon fluxes and growth front that is straight at the moir\'{e} pattern scale on Ir(111) surface. As illustrated in Fig.\ref{fig:180ts}(a), kMC is performed on a huge lattice containing $L_{x}\times L_{y}=80\times48$ moir\'{e} patterns each of which consists of $10\times20$ growth sites with an initial grown graphene ribbon of size $2*48$ moir\'{e} patterns. The rate constants are $k_a=2.18\times10^{8}$, $k_b=2.18$ and $k_c=2.455\times10^{-4}$ for monomer carbon concentration $n_1=0.01$ monolayer and $T=1170K$ by taking $f_{0,c}$ to be the unit which leads to the magnitude of $f_{0,a}$ is about $10^3$ and $f_{0,b}\approx 1$. In practice, we find that the growth kinetics is very robust to the exact values of these rate constants due to the large discrepancies between them.

In this case, the rate-dominating step is the attachment of $C_5$ on the D-sites of open-type. Therefore, the growth rate of the graphene area $dA/dt$ is proportional to the number of configurations on the growth front that allow the rate-dominating event happening (denoted here by $n_{rdc}$). In Fig.\ref{fig:180ts}(b), we plot $A$ (normalized by the area of initial grown ribbon) as a function of time $t$, where a perfect linear dependence is observed, indicating $n_{rdc}$ are nearly constant along the growth front. This is reasonable since $n_{rdc}$ should be proportional to the length of the growth front, here is just $L_y$ which is a constant. In the inset of Fig.\ref{fig:180ts}(b), time averaged area-growth rate $\left\langle dA/dt\right\rangle $ is plot as a function of the monomer concentration $n_1$. The curve obtained can be fitted by $\left\langle dA/dt\right\rangle =a+bn_{1}^{\gamma}$ with an exponent $\gamma\approx5$, which is well consistent with observations in experiments\cite{NJP08093026} and in previous simulation\cite{JAC12006045}. Thus we believe that the minimal model is able to capture the essential physical picture regarding the growth kinetics on Ir surface with graphene-substrate lattice mismatch.

\begin{figure}[h]
\begin{centering}
\includegraphics[width=0.9\columnwidth]{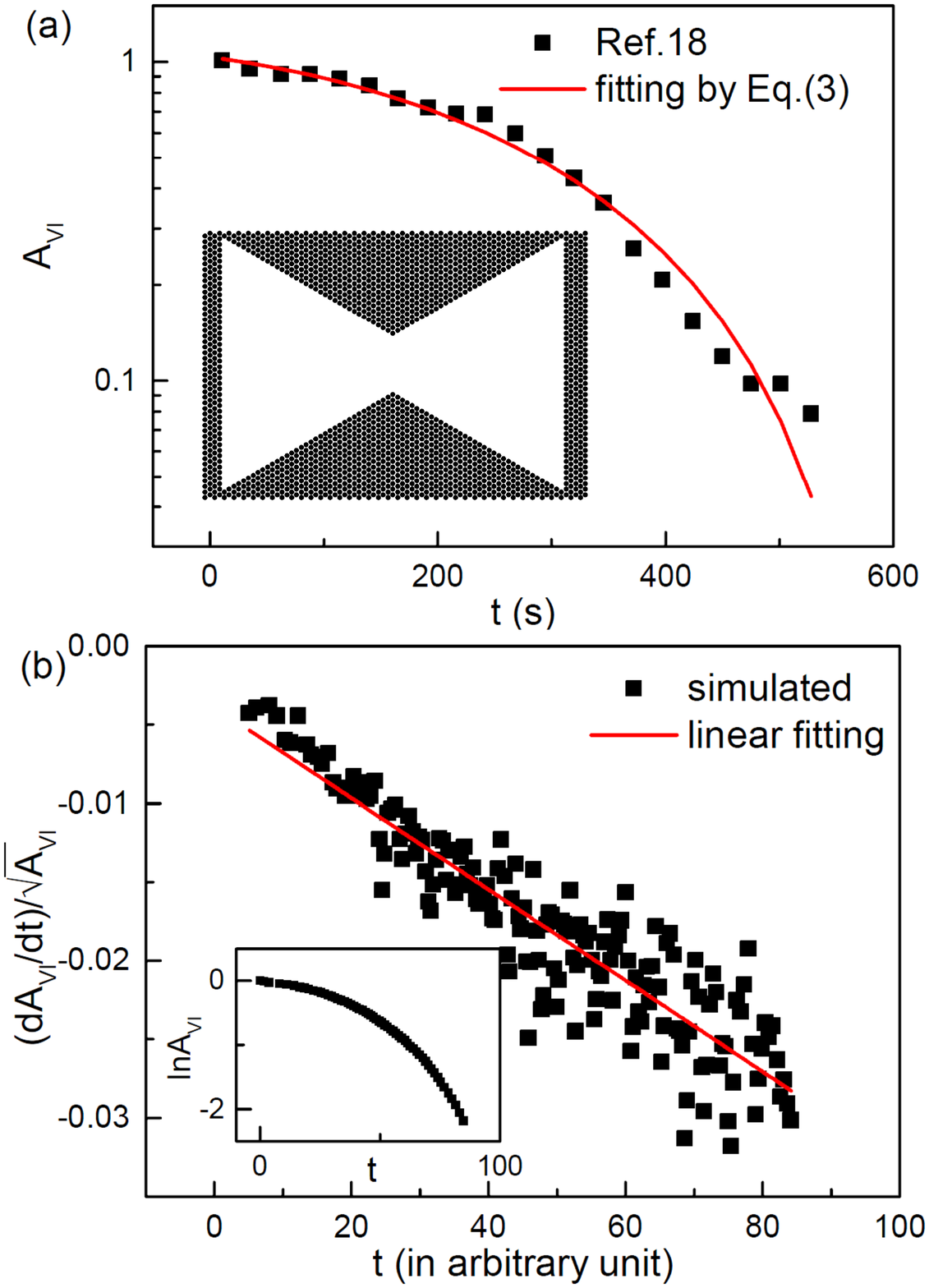}
\par\end{centering}
\protect\caption{(Color online) (a) The area $A$ obtained experimentally by \citeauthor{ACSNano13007028}\cite{ACSNano13007028} (black squares) as a function of time $t$ for graphene growth over VI, which can be well fitted by Eq.(\ref{eq:area}) (the red line) derived from the minimal kMC simulation. The inset shows the initial shape of VI at the moir\'{e} pattern scale. (b) Linear dependence of the growth rate $dA_{VI}/dt$ normalized by $\sqrt{A_{VI}}$ on $t$. The inset shows the simulated time-resolved $A_{VI}$ of VI. }
\label{fig:ts}
\end{figure}

\section{Application on Growth over Vacancy Island}
Now we use our approach to investigate the growth behavior with complex front shapes. As a typical example, we consider a VI as depicted in the inset of Fig.\ref{fig:ts}(a) wherein several fragments merge into on piece of grown graphene. The growth kinetics over such a specific shape has been investigated in some details experimentally\cite{ACSNano13007028}. It has been found that the area $A_{VI}$ (the area of the empty region inside the graphene flake) normalized by its initial value exhibits a complicated dependence on time, which is surely non-exponential as discussed by \citeauthor{ACSNano13007028}\cite{ACSNano13007028}, while the underlying mechanism is yet not clear. Interestingly, using our approach with all the parameters same as above mentioned, we can obtained the time-dependence of $A_{VI}$ (the inset of Fig.\ref{fig:ts}(b)) which is very similar in tendency with that observed in the experiments (scatters in Fig.\ref{fig:ts}(a)). In addition, our model makes it convenient to investigate in detail how the boundary of the VI change with time. As discussed above, for a regular front as shown in Fig.\ref{fig:ts}(a), one can expect that $dA_{VI}/dt\propto\sqrt{A_{VI}}$ since $n_{rdc}\propto\sqrt{A_{VI}}$. Nevertheless, for irregular boundary here, we find that $dA_{VI}/dt$ is not proportional to $\sqrt{A_{VI}}$ at all. Rather surprisingly, $(dA_{VI}/dt)/\sqrt{A_{VI}}$ in nearly linear with time $t$, as drawn in Fig.\ref{fig:ts}(b). One may then write
\begin{equation}
\frac{dA_{VI}(t)}{dt}=\sqrt{A_{VI}(t)}(mt+n),\label{eq:rate}
\end{equation}
which gives
\begin{equation}
A_{VI}(t)=(\frac{1}{2}mt^{2}+nt+p)^{2}\label{eq:area}
\end{equation}
where, $m$, $n$ and $p$ are certain fitting parameters. Remarkably, we find that such a function can fit the experimental data quantitatively well as demonstrated in Fig.\ref{fig:ts}(a) with $m=-3.74\times10^{-6}$, $n=-5.47\times10^{-4}$ and $p=1.02$. Therefore, we believe that Eq.(\ref{eq:rate}) and (\ref{eq:area}) have captured some basic physical features regarding the growth of graphene over the VI.

\begin{figure}[h]
\begin{centering}
\includegraphics[width=1\columnwidth]{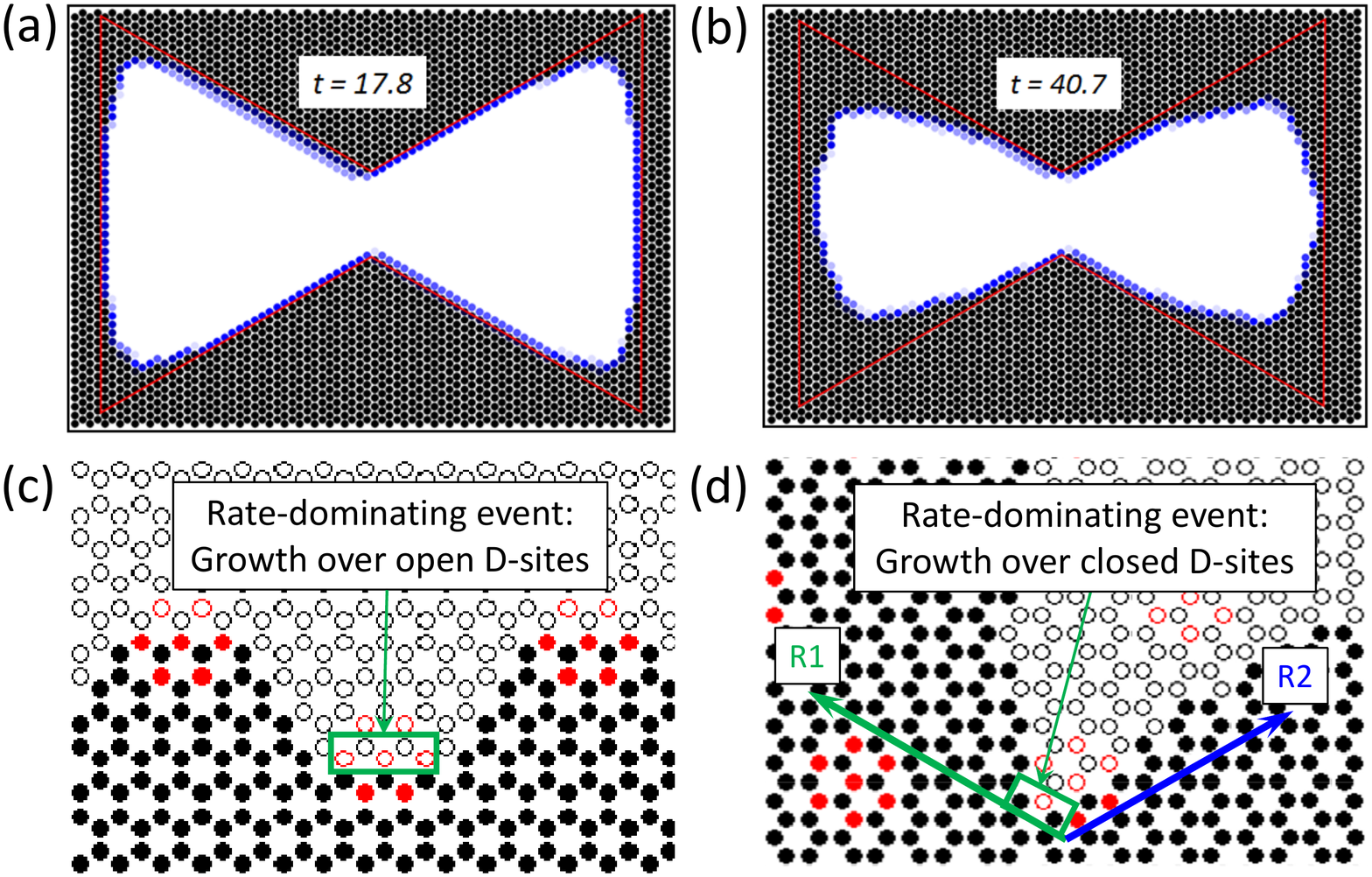}
\par\end{centering}
\protect\caption{(Color online) Typical kMC snapshots for growth over VI at (a) $t=17.8$ (in arbitrary unit) and (b) $t=40.7$ at the moir\'{e} pattern scale. Red lines illustrate the initial shape of VI. (c) and (d) are typical zoom-in atomic-level structures for a straight segment and a concave one, respectively, where each circle refers to a growth site and E-sites/D-sites is colored by black/red. The green rectangles enclose typical configurations of rate-dominating events for growth over D-sites. In (d), D-site is actually of closed-type along a given zigzag front indicated by $R_{1}$, although along $R_2$ it is of open-type.}
\label{fig:snapshot}
\end{figure}

To obtain a deeper understanding about the interesting growth behavior of VI, we try to find hints from spatial information contained in snapshots of the surface during growth process, which is one of the major advantages of kMC simulations. Two snapshots of VI are shown in Fig.\ref{fig:snapshot}(a) and (b) for $t=17.8$ and $t=40.7$  (in arbitrary unit), respectively, where the initial shape is outlined by the red lines. Compared to the initial profile, it can be observed that the growth process mainly takes place at the \textit{concave} segments in the corner parts, while the convex parts in the middle remain nearly unchanged. This indicates that the number of relevant configuration $n_{rdc}$ should be proportional to the length of concave segments within the growth front, $L_{c}$, rather than to the total length of growth front $L$. Comparing the snapshots for $t=17.8$ and $t=40.7$, one can see that the ratio $L_c/L$ increases with time. To be in accordance with Eq.(\ref{eq:rate}), one expects that $L_{c}/L\propto t$ assuming that $L\propto \sqrt{A_{VI}}$. Nevertheless, the reason of this particular linear dependence with time is still open to us at the current stage.

Now the key point turns to why the effective number of configurations $n_{rdc}$ for growth is proportional to the length of concave front segments. To this end, we depict typical zoom-in atomic-level structures for a straight segment and a concave one in Fig.\ref{fig:snapshot}(c) and (d), respectively. Clearly, for the straight segment, the D-site is of open-type, such that the rate-dominating step still has to be $C_{5}$-attachment. However, for the concave segment formed within the corner environment, the D-site is actually of closed-type along a given zigzag front indicated by $R_{1}$, although along $R_2$ it is of open-type. Such a dynamic change from open-type to closed-type of the D-site due to the local shape of the front leads to a change of the rate-dominating event from $C_5$-attachment to $C_1$-attachment. In other words, the rate-dominating event (growth over D-sites) is heterogeneous for different parts of the growth front, i.e., $C_{1}$ attachment for concave segment, and $C_{5}$ attachment for others. Consequently, the overall growth rate of graphene is determined by the length of concave segments. Noted that the energetic parameters for all growth events are independent on the front shape, the heterogeneity of rate-dominating events is solely geometry effects.

\section{Conclusion}
In summary, we have built up a minimal kinetic Monte Carlo model to study large-scale growth kinetics of graphene on active catalyst metal surfaces with graphene/substrate lattice mismatch. Using kinetic parameters extracted from detailed atomic calculations, our model successfully reproduced the time-resolved graphene growth over vacancy island observed in experiments, and revealed that the interesting dependence of the area on time is resulted from the heterogeneous rate-dominating events for graphene growth at different growth front segments, which was shown to be mainly determined by geometry. Since the proposed minimal kMC model (or perhaps its modified version) provides a powerful theoretical tool for investigation of large-scale growth kinetics of graphene, and our findings take an important step forward to the underlying growth mechanism, we hope that the present study could open new perspectives on theoretical studies and motivate more atomic-resolution imaging experiments.

\begin{acknowledgments}
This work is supported by National Basic Research Program of China (2013CB834606), by National Science Foundation of China(21125313, 21473165, 21403204), and by the Fundamental Research Funds for the Central Universities (WK2060030018,2340000034).
\end{acknowledgments}

\end{document}